\documentclass[12pt]{article}

\usepackage[sectionbib]{natbib}
\usepackage{array,epsfig,fancyhdr,rotating}
\usepackage[doublespacing]{setspace}
\usepackage[dvipdfm]{hyperref}

\usepackage{amsmath}
\usepackage{amssymb}
\usepackage{amsfonts}
\usepackage{multirow}
\usepackage{amsthm}

\usepackage{color}

\usepackage{algorithm}
\usepackage{algorithmic}

\begin{document}

\title{Unbiased local solutions of partial differential equations via the Feynman-Kac Identities}

\author{Jake Carson$^1$\footnote{Email: Jake.Carson@warwick.ac.uk}, Murray Pollock$^1$ and Mark Girolami$^{12}$}

\date{ }

\maketitle

\begin{center}
{\small
 $^1$ University of Warwick, Department of Statistics, UK. \\
 $^2$ The Alan Turing Institute for Data Science, British Library, UK. \\}
\end{center}

\begin{abstract}

The Feynman-Kac formulae (FKF) express local solutions of partial differential equations (PDEs) as expectations with respect to some complementary stochastic differential equation (SDE). Repeatedly sampling paths from the complementary SDE enables the construction of Monte Carlo estimates of local solutions, which are more naturally suited to statistical inference than the numerical approximations obtained via finite difference and finite element methods. Until recently, simulating from the complementary SDE would have required the use of a discrete-time approximation, leading to biased estimates. In this paper we utilize recent developments in two areas to demonstrate that it is now possible to obtain unbiased solutions for a wide range of PDE models via the FKF. The first is the development of algorithms that simulate diffusion paths exactly (without discretization error), and so make it possible to obtain Monte Carlo estimates of the FKF directly. The second is the development of debiasing methods for SDEs, enabling the construction of unbiased estimates from a sequence of biased estimates.

\end{abstract}

Keywords: Debiasing; Exact algorithms; Feynman Kac formula; Multilevel Monte Carlo; Partial differential equations; Stochastic differential equations.

\noindent {\bf 1. Introduction}

Partial differential equations (PDEs) describe many natural and physical phenomena, and have long been the foundation of mathematical models used for the study of complex behaviour. 
These can be considered forward models, in the sense that for some given set of model parameters they generate a set of possible observations. 
In many situations the goal is to use observations to estimate the model parameters, and in so doing, better understand the underlying processes of the system. 
This is known as the inverse problem, which is detailed in, for example, \citet{Stuart2010}. 
In most cases analytical solutions to PDEs of interest are unavailable, and so solutions are usually obtained using a numerical approximation. 
Approaches such as finite element and finite difference methods provide discrete space-time approximations to the solutions at specified grid-points over regions of interest. 
These approximate solutions contain systematic errors induced by the discretization of the PDEs. 
Whilst the convergence properties of these methods are usually known as the grid is refined, the magnitude of the error is almost always unavailable. 
This is problematic when performing statistical inference in both the forward and inverse problems, as it is often not possible to construct valid confidence intervals, or Bayesian credible intervals.

\lhead[\footnotesize\thepage\fancyplain{}\leftmark]{}\rhead[]{\fancyplain{}\rightmark\footnotesize\thepage}

An alternative approach is to use the Feynman-Kac formulae (FKF) \citep{Kac1951}, which describe the relationship between the local solutions to certain classes of PDEs and the expectations of initial and boundary functions with respect to the law of complementary stochastic differential equations (SDEs). 
The FKF approach can only be applied to certain classes of PDEs, but for such cases this approach may have a number of advantages when performing statistical inference. 
Computationally, the FKF approach may be more efficient in high dimensional problems as the number of grid-points in global solvers increases exponentially with the number of dimensions, whereas the cost of simulating from the complementary SDEs grows sub-exponentially. 
An additional advantage in the inverse problem is that the number of observations will typically be far fewer than the number of grid points required in a finite differences or finite elements solution, and so obtaining only the required local solutions may further reduce the computational cost. 
The estimates of each local solution are also independent, and so can be run in parallel, drastically reducing wall-clock times. 
Finally, the Monte Carlo error can be studied using, for example, the central limit theorem, offering a more natural approach for statistical inference than the error bounds of finite difference and finite element methods. 
The unbiased local solutions can then be embedded in either a Bayesian or maximum likelihood framework in order to study the inverse problem.

Analytical solutions to the FKF are rarely available, but it is usually possible to obtain Monte Carlo estimates by sampling a finite number of random paths from the complementary SDE.
Owing to the difficulty of simulating from SDEs, existing applications of the FKF have suffered various limitations.
The random walk on spheres (WOS) algorithm \citep{Muller1956}, for example, was proposed to obtain Monte Carlo solutions to the Dirichlet problem of Laplace's equation.
The WOS algorithm has since been extended (see for example \citep{Hwang2001,Hwang2003}), but is still restricted to a narrow class of models, in particular requiring that the drift and diffusion functions of the complementary SDE are constant.
In all cases solutions are approximate as simulated paths are terminated in a region near the boundary, rather than on the boundary itself.
A more recent modification is the random walk on rectangles (WOR) algorithm \citep{Deaconu2006}.
If the domain is a polytope then the WOR algorithm can be used to simulate the absorption time exactly, and so the only error in the estimate of the FKF is that associated with the Monte Carlo approximation.
In many situations the drift and/or volatility functions will be non-constant, preventing the application of the WOS and WOR algorithms.
In such cases estimates of the FKF can be obtained by using discrete time approximations of the SDE, such as the Euler-Maruyama approximation \citep{Kloeden1992}, for example.
This approach was used in \citet{Herbei2014} to perform inference on an ocean-circulation model.
However, such numerical schemes lead to biased estimates of the FKF due to discretization errors.

In this article we demonstrate how advancements in two areas can be used to obtain unbiased estimates of the FKF for a wider range of models.
The first is the recent development of exact algorithms (EA).
These are rejection algorithms that use Brownian motion proposals to obtain finite representations of SDEs of interest, and are exact in the sense that they do not contain discretization errors.
In particular we extend the localisation approach of \cite{Chen2013} to multiple dimensions, and design an adaptive scheme to simulate the exact first passage time from a hyperrectangle, allowing us to directly obtain Monte Carlo estimates of solutions to certain Dirichlet problems.
The EA approach imposes restrictions on the drift and volatility functions of the complementary SDE, and so is not always available.
The second is the development of debiasing techniques for expectations with respect to SDEs.
Debiasing techniques construct unbiased estimates using a weighted sum of biased estimates that are obtained from numerical approximations.
Since debiasing only requires that we are able to simulate from the SDE using approximate (but convergent) schemes, it has fewer restrictions than EA, and so can be more generally applied, albeit at the cost of larger work-variance products.


The layout of this paper is as follows. 
In Section 2 we discuss the FKF, demonstrating how to represent local solutions of PDEs as expectations with respect to the law of complementary SDEs. 
In Section 3 we discuss how exact algorithms can be used to obtain unbiased local solutions to PDEs via exactly simulating from the complementary SDEs. 
In Section 4 we discuss how unbiased estimates can be obtained using discrete-time approximations of the complementary SDEs via debiasing. 
In Section 5 we give some numerical examples in order to compare the two approaches. 
In Section 6 we conclude and discuss future directions for research.\\

\noindent {\bf 2. Feynman-Kac Formulae}

The FKF relate solutions of a class of deterministic second order PDEs with expectations of initial and boundary functions with respect to the law of complementary SDEs. An introduction is available in \citet{Oksendal2000}, and covered in more detail in \citet{Pardoux2014}. Consider the case where $u(\pmb{x},t) \in C^{2,1} \left(\mathbb{R}^d \times [0,\infty) \right)$ is the solution of a system described by the PDE
\begin{equation}
\label{E:PDE}
\frac{\partial u}{ \partial t} = \sum_{i,j=1}^{d} \frac{1}{2} a_{ij}(\pmb{x},t) \frac{\partial^2 u}{\partial x_i \partial x_j} + \sum_{i=1}^{d} b_i (\pmb{x},t) \frac{\partial u}{\partial x_i} - c(\pmb{x},t) u, \hspace{20pt} \pmb{x} \in \mathbb{R}^{d}, \; t>0.
\end{equation}
\noindent with the initial condition $u(\pmb{x},0) = f(\pmb{x})$. This is known as the initial value problem. We assume that $ c(\pmb{x},t)$ is positive and bounded above, and that the coefficients in (\ref{E:PDE}) satisfy the necessary conditions to ensure the existence of a unique solution (see \citet{Evans1998}, for example). The FKF allow us to represent the local solution of the PDE, $u(\pmb{x},t)$, as the expectation of the initial function, $f(\pmb{x})$, with respect to the law of a complementary SDE satisfying
\begin{equation}
\label{E:CSDE}
{\rm d}\pmb{X}_s^x = b(\pmb{X}_s^x, t - s) {\rm d}s + \pmb{\sigma} (\pmb{X}_s^x, t - s) {\rm d}\pmb{W}_s,
\end{equation}
\noindent where $\pmb{W}_s \in \mathbb{R}^d$ is a vector of independent Brownian motions at time $s$, $\pmb{X}_s^x  \in \mathbb{R}^d$ is the state of the SDE at time $s$ and initialised at $\pmb{x}$, i.e. $\pmb{X}_0^x = \pmb{x}$, and $\pmb{\sigma} $ satisfies  $\pmb{a}_{ij}(\pmb{x},s) = \sum_k \sigma_{ik}(\pmb{x},s) \sigma_{kj}(\pmb{x},s)$. In this specific case,
\begin{equation}
\label{E:FK}
u(\pmb{x},t) = E \left\{ f \left(\pmb{X}_t^x \right) \exp \left(- \int_0^t c \left(\pmb{X}_s^x, t-s \right) ds \right) \right\}.
\end{equation}
\noindent Note that we have further assumed that $\pmb{b}(\pmb{x},s)$ and $\pmb{\sigma}(\pmb{x},s)$ are sufficiently regular to ensure a unique, non-explosive, weak solution to (\ref{E:CSDE}) (see \citet{Karatzas1991}, for example).

In nontrivial cases (\ref{E:FK}) will need to be estimated using Monte Carlo methods. In comparison to deterministic solvers we have replaced the deterministic discretization error with Monte Carlo error, and so the error analysis of (\ref{E:FK}) can be performed with the Central Limit Theorem (CLT), for example \citep{Graham1996}. This probabilistic approach is more naturally suited for performing statistical inference.

\noindent {\bf 2.1. Bounded Domains}

In many situations the PDE is bounded on the domain $\Omega$. Common boundary conditions are Dirichlet boundary conditions, defined by 
\begin{equation*}
u(\pmb{x},t) = g(\pmb{x},t) \hspace{30pt} \pmb{x} \in \partial \Omega,
\end{equation*}
\noindent where $\partial \Omega$ is the boundary of the domain, and Neumann boundary conditions, defined by 
\begin{equation*}
\frac{\partial u(\pmb{x},t)}{\partial \pmb{n}} = h(\pmb{x},t) \hspace{30pt} \pmb{x} \in \partial \Omega,
\end{equation*}
\noindent where $\pmb{n}$ denotes the normal to the boundary. The FKF can be modified to include such boundary conditions by modifying the law of the complementary SDE \citep{Pardoux2014,Skorokhod1962}. For example, consider a system on the domain $\Omega$ with Dirichlet boundary conditions. We now have a PDE of the form
\begin{equation*}
\frac{\partial u}{ \partial t} = \sum_{i,j=1}^{d} \frac{1}{2} a_{ij}(\pmb{x},t) \frac{\partial^2 u}{\partial x_i \partial x_j} + \sum_{i=1}^{d} b_i (\pmb{x},t) \frac{\partial u}{\partial x_i} - c(\pmb{x},t) u, \hspace{20pt} \pmb{x} \in \Omega, t>0,
\end{equation*}
\noindent with initial and boundary conditions
\begin{equation*}
\begin{aligned}
u(\pmb{x},0) & = f(\pmb{x}), & \pmb{x} & \in \Omega,\\
u(\pmb{x},t) & = g(\pmb{x}), & \pmb{x} & \in \partial \Omega, \; t>0.
\end{aligned}
\end{equation*}
\noindent The expectation in the FKF is now with respect to the law of an absorbed Brownian motion. Define a stopping time,
\begin{equation*}
\gamma_\Omega^{x,t} = \inf \left\{ s \geq 0 \mid \pmb{X}_s^x \in \mathbb{R}^d \backslash \Omega, \right\}
\end{equation*}
\noindent which is the first time at which the stochastic process hits the boundary, $\partial \Omega$, and is absorbed. Define a second stopping time, $\hat{t} = \min(\gamma_\Omega^{x,t},t)$, which returns $t$ if the process does not hit the boundary before time $t$, and returns the first time at which the stochastic process hits the boundary otherwise. For convenience, define:
\begin{equation*}
k(\pmb{x}) = \left\{ \begin{aligned} & f(\pmb{x}), \hspace{5pt} \mbox{if } \hat{t} = t, \\  
& g(\pmb{x}), \hspace{5pt} \mbox{if } \hat{t} < t, \end{aligned} \right.
\end{equation*}
\noindent In this case the FKF takes the form
\begin{equation}
\label{E:FKD}
u(\pmb{x},t) = E \left\{ k \left(\pmb{X}_{\hat{t}}^x \right) \exp \left(- \int_0^{\hat{t}} c \left(\pmb{X}_{s}^x, t-s \right) ds \right) \right\}.
\end{equation}
\noindent In the case of Neumann boundary conditions a similar modification can be used, in which the Brownian motion is reflected rather than absorbed \citep{Pardoux2014}. 
\\

\noindent {\bf 3. Obtaining FKF estimates via path space importance sampling}

The law of the complementary SDE is rarely available in closed-form. 
We must resort to numerical simulation of the SDE in order to obtain a Monte Carlo estimate of the integrals in (\ref{E:FK}) and (\ref{E:FKD}). 
Until recently this would require the use of a discrete time approximation, such as the Euler-Maruyama method, leading to bias in the FKF estimates.
The development of path space importance sampling algorithms (commonly referred to as exact algorithms, or EA) have now made exact  simulation (without discretisation error) from a wide range of SDEs possible.
In turn, by constructing the proposal in a particular manner, it is possible to obtain unbiased estimates of local solutions of parabolic/elliptic PDEs with Dirichlet boundary conditions on polytopal domains.
These developments are described in the remainder of this section.
For notational convenience we will limit the initial discussion to time-homogenous univariate diffusions, and cover relevant extensions at the end of the section. 
In addition to the assumptions given in Section 2, for the remainder of the section we assume:
\begin{enumerate}
\item Continuity assumption: The drift coefficient $b(x,s) \in C^1$, and the volatility coefficient is strictly positive and $\sigma(x,s) \in C^2$.
\item Growth bound assumption: There exists some constant $K>0$ such that $\left| b(x) \right|+ \left| \sigma (x) \right| \leq K(1+\left| x \right|^2) \mbox{ for all } x \in \mathbb{R}$.
\end{enumerate}
\noindent These assumptions are sufficient for the EA methodology presented here. In particular, for application of EA it is typically necessary for the target SDE to have unit volatility, ensuring that standard Brownian motion provides an equivalent measure. Assumptions 1 and 2 permit the use of the Lamperti transform \citep{AitSahalia2002}, as follows: From the complementary SDE, construct a new process 
\begin{equation*}
{\rm d}Y_s^y = \eta(X_s^x) = \int_z^{X_s^x}{ \frac{1}{\sigma(v)} {\rm d}v },
\end{equation*}
\noindent so that transformed SDE then takes the form
\begin{equation*}
\label{E:TSDE}
{\rm d}Y_s^y = \alpha(Y_s^y) {\rm d}s + {\rm d}W_s,
\end{equation*}
\noindent where superscript $y$ indicates that $Y^y_0 = y$, and
\begin{equation*}
\alpha(v) = \frac{b\left((\eta^{-1}(v)\right)}{\sigma \left((\eta^{-1}(v)\right)} - \frac{1}{2}\sigma'\left(\eta^{-1}(v) \right).
\end{equation*}

EA utilises rejection sampling \citep{VonNeumann1951} to sample a {\em skeleton} of an SDE of interest. A skeleton is an exact (without discretisation error) sample of the SDE at a finite number of time points, which additionally contains information about the time interval of the path, $[0,t]$, and path space information that ensures the sample path is almost surely constrained to some compact interval \citep{Pollock2014}. Denoting the target measure (the law of the SDE) as $\mathbb{T}_{[0,t]}$, skeletons are obtained via sampling from an equivalent proposal measure, $\mathbb{P}_{[0,t]}$, which is often taken to be standard Brownian motion. The probability of accepting proposed paths is proportional to the Radon-Nikod\'{y}m derivative of $\mathbb{T}_{[0,t]}$ with respect to $\mathbb{P}_{[0,t]}$ \citep{Oksendal2000}.

In order to propose sample paths, we utilise the methodology developed in \citet{Chen2013}, which can be considered as a localised EA (LEA) approach.
The localised approach partitions the simulation of a sample path by first fixing an interval around the initial value and then probabilistically samples an exit time and location for the proposed path.
The simulated exit point then becomes the initial value in the next iteration of the algorithm, and this continues until the entire path is sampled.
More formally, consider the sequence of first passage times 
\begin{equation*}
\zeta^{(i)} = \inf \left(s \geq 0 : \left| Y_{\zeta^{(i)}} - Y_{\zeta^{(i-1)}} \right| > \theta \right), \: \: i \geq 1
\end{equation*}
\noindent for some user defined constant $\theta$, which gives the half-width of the interval. Furthermore denote $(\zeta^{(0)}=0, Y_{\zeta^{(0)}} = y)$ as the start point of the (transformed) complementary SDE. In LEA the pairs $(\zeta^{(i)}, Y_{\zeta^{(i)}})$ are sequentially simulated until the target time, $t$, has been surpassed. This requires the simulation of two random variables, $\Delta \zeta^{(i)} = \zeta^{(i)} - \zeta^{(i-1)}$ and $\Delta Y^{(i)} = Y_{\zeta^{(i)}} - Y_{\zeta^{(i-1)}}$, which bound the path of $Y_s$.

Brownian motion is used as the proposal distribution, requiring that we are able to sample $\Delta \varsigma^{(i)} = \varsigma^{(i)} - \varsigma^{(i-1)}$ and $\Delta W^{(i)} = W_{\zeta^{(i)}} - W_{\zeta^{(i-1)}}$, where $W_s$ is a standard Brownian motion, and 
\begin{equation*}
\label{E:BMFTP}
\varsigma^{(i)} = \inf (s \geq 0 : \left| W_{\varsigma^{(i)}} - W_{\varsigma^{(i-1)}} \right| > \theta ), \: \: i \geq 1,
\end{equation*}
with $\left(\varsigma^{(0)}=0, W_{\varsigma^{(0)}} = y \right)$. The first passage times of Brownian motion from a symmetric interval can be efficiently sampled using the rejection algorithm developed in \citet{Devroye2009}, which has a rejection rate of less than $0.001$.
Once the first passage time is sampled, the symmetric property of Brownian motion means that 
\begin{equation*}
P(\Delta W^{(i)} = \theta) = P(\Delta W^{(i)} = -\theta) = \frac{1}{2}.
\end{equation*}
\noindent The value of $\theta$ does not need to remain constant, and we could instead consider selecting a sequence of constants, $\theta^{(i)}$. In particular, when the SDE is bounded as in the Dirichlet problem, then we can exactly simulate the absorption time of the path by adaptively choosing $\theta^{(i)}$ so that one side of the interval coincides with the nearest boundary. It is also beneficial to include an upper limit, $\tilde{\theta} = \sqrt{t-\varsigma^{(i-1)}}$, where $t$ is the target time, in order to avoid the extra computational expense incurred when sampled first passage times are significantly larger than the target time. When a sampled $\varsigma^{(i)}$ exceeds the target time, the terminal value, $W_t$, can be sampled using the law of a Bessel bridge with minimum (or maximum) $W_{\varsigma^{(i)}}$.

Denoting $\mathbb{Q}_{[\varsigma^{(i-1)},\varsigma^{(i)} \wedge t]}$ as the law of the transformed SDE, and $\mathbb{W}_{[\varsigma^{(i-1)},\varsigma^{(i)} \wedge t]}$ as the law of a standard Brownian motion, the Radon-Nikod\'{y}m derivative of $\mathbb{Q}_{[\varsigma^{(i-1)},\varsigma^{(i)} \wedge t]}$ with respect to $\mathbb{W}_{[\varsigma^{(i-1)},\varsigma^{(i)} \wedge t]}$ is 
\begin{equation}
\label{E:ChenRND}
\frac{{\rm d} \mathbb{Q}_{[\varsigma^{(i-1)},\varsigma^{(i)} \wedge t]}}{{\rm d} \mathbb{W}_{[\varsigma^{(i-1)},\varsigma^{(i)} \wedge t]}} (Y) \propto \exp \left\lbrace A(W_{\varsigma^{(i)} \wedge t}) - \int_{\varsigma^{(i-1)}}^{\varsigma^{(i)} \wedge t} \phi(Y_s) {\rm d}s  \right\rbrace ,
\end{equation}
\noindent where $A(v)=\int_0^v{\alpha(z)dz}$, and $\phi(v)=\frac{1}{2}\left(\alpha(v)^2+\alpha'(v)\right)$. It was noted in \citet{Chen2013} that the continuity of $\phi(v)$ ensures that the Radon-Nikod\'{y}m derivative is bounded, allowing the construction of a valid acceptance probability proportional to (\ref{E:ChenRND}). 
This requires calculating an upper bound for $\exp (A(v))$, and upper and lower bounds for $\phi(v)$ over the current interval. For the Dirichlet problem it is possible to obtain bounds over the entire domain, which can then be applied to every interval. Whether this is a more computationally efficient approach than evaluating tighter bounds in each iteration will be problem dependent. Whilst it is not possible to compute the acceptance probability directly as it requires the full sample path, an unbiased acceptance probability can be obtained using Poisson thinning \citep{Lewis1979}, which requires only a finite dimensional representation of the sample path \citep{Beskos2006,Chen2013}. 

Accepted simulations from EA can be used to obtain Monte Carlo estimates of the FKF by using the inverse of the Lamperti transform. Since the function $c(x,t)$ is required to be bounded, say by $L_c$ and $M_c$, the path weighting term in (\ref{E:FK}) and (\ref{E:FKD}) can also be estimated using Poisson thinning. A naive Monte Carlo estimate is then given by
\begin{equation*}
\hat{u} \left(\pmb{x}, t \right) = \frac{1}{N} \sum_{i=1}^{N} \hat{u}^{(i)},
\end{equation*}
\noindent where each $\hat{u}^{(i)}$ are of the form
\begin{equation*}
\hat{u}^{(i)} = \kappa(\eta^{-1}(Y_{\hat{t}})) \prod_{j=1}^{N_p}{\frac{ \left( M_c - c(\eta^{-1}(Y_{r_j}),t-r_j) \right)}{M_c-L_c}} \cdot \exp \left\lbrace -L_c \hat{t} \right\rbrace,
\end{equation*}
\noindent and $r_j, \; j=1,...,N_p$ are times sampled from a Poisson process with intensity $M_c-L_c$ over $[0,\hat{t}]$. The variance of this estimator can be reduced by sampling $N_p$ from alternative distributions, as discussed in Section 4 of \citet{Fearnhead2008}. Note also that we can simulate values of $\hat{u}^{(i)}$ from the proposal measure by using LEA with the modified function $\tilde{\phi}(u) = \phi(u) + c(u)$, rather than use a two-step process.

We now consider extending LEA to multiple dimensions.
Whilst the Lamperti transformation is usually available for univariate diffusions, it is rarely available in the multivariate case \citep{AitSahalia2008}. This is the first major restriction on the applicability of exact algorithms to multivariate diffusions. In the remainder of this article we assume that such a transformation is possible, and denote the transformed process
\begin{equation*}
\label{E:MVTSDE}
{\rm d}\pmb{Y}_s^y = \pmb{\alpha}(\pmb{Y}_s^y) {\rm d}s + {\rm d}\pmb{W}_s.
\end{equation*}
For the proposal we use $d$ independent Brownian motions, although we could in principle use correlated Brownian motions if required.
When the domain of the PDE of interest forms a hyperrectangle we adaptively choose the bounding interval lengths in each path segment, $\pmb{\theta}^{(i)} = (\theta^{(i)}_{(1)},...,\theta^{(i)}_{(d)})$, to coincide with the nearest boundary in each dimension. 
The methodology presented here can be adapted to any polytopal domain with appropriate rotation of the coordinate axes and selection of bounding intervals.
As this algorithm shares many features with the random walk on rectangles algorithm, insight into selecting bounding intervals on polytopal domains can be found in \citet{Deaconu2006}.
Applying EA to other domains is considered in Section 6.
We require the multivariate analogues of the functions $\phi(v)$ and $A(v)$, which are derived in \citet{Beskos2006b}. For $\phi(v)$ this is $\phi(\pmb{v}) = \frac{1}{2} \left( \| \pmb{\alpha}(\pmb{v}) \|^2 + \nabla \cdot \pmb{\alpha}(\pmb{v}) \right)$, and $A(\pmb{v})$ is defined through the relation $\nabla A(\pmb{v}) = \pmb{\alpha}(\pmb{v})$. Requiring the latter is the second major restriction regarding applicability of EA, as no such potential function, $A(\pmb{v})$, exists for a large number of SDEs. Finally, for time-inhomogenous diffusions define $\phi(\pmb{u}, s) = \frac{1}{2} \left( \| \pmb{\alpha}(\pmb{u}, s) \|^2 + \nabla \cdot \pmb{\alpha}(\pmb{u}, s) + 2 \frac{\partial}{\partial s} A(\pmb{u}, s) \right)$, and $A(\pmb{u}, s)$ through the relation $\nabla A(\pmb{u}, s) = \pmb{\alpha}(\pmb{u}, s) $. \\

\noindent {\bf 4. Obtaining FKF Estimates via Debiasing}

For applications for which EA is not viable, it is possible to resort to numerical simulation of the SDE in order to obtain Monte Carlo estimates of (\ref{E:FK}) and (\ref{E:FKD}). Take for example the Euler-Maruyama scheme, which divides the time domain $[0,t]$ in to $M$ equal intervals of size $h=\frac{t}{M}$, and then simulates a path by iteratively drawing values from the multivariate Gaussian distribution
\begin{equation*}
\label{E:TD}
\tilde{\pmb{X}}^x_{(m+1)h} \sim \mathcal{N}_d\left( \; \tilde{\pmb{X}}^x_{mh} + \pmb{b}(\tilde{\pmb{X}}^x_{mh},t-mh) h, \; \pmb{a}(\tilde{\pmb{X}}^x_{mh},t - mh) h \; \right),
\end{equation*}
\noindent for $m=0,...,M-1$, subject to the condition $\tilde{\pmb{X}}^x_0 = \pmb{x}$. The two arguments are the mean and variance of the distribution respectively. Off-grid values can be interpolated using, for example, a linear or piecewise constant process. An approximate Monte-Carlo approximation of (\ref{E:FK}) is then obtained by sampling
\begin{equation*}
\label{E:FKEulSim}
\tilde{u}_h^{(i)} = f(\tilde{\pmb{X}}_{Mh}^{x}) \exp \left(- h \sum_{m=0}^{M-1} c\left(\tilde{\pmb{X}}_{mh}^{x},t - m h \right) \right),
\end{equation*}
\noindent for $i=1,...,N$, and averaging
\begin{equation}
\label{E:FKEulMC}
\tilde{u}_h (\pmb{x}, t) = \frac{1}{N} \sum_{i=1}^N \tilde{u}_h^{(i)}.
\end{equation}
\noindent An error analysis of (\ref{E:FKEulMC}) is available in \citet{Graham1996}. For an approximate Monte-Carlo approximation of (\ref{E:FKD}) the simulation is stopped as soon as the path exits the domain. 
The estimate $\tilde{u}_h (\pmb{x}, t)$ is biased for any $h > 0$, but $E\left\lbrace \tilde{u}_h (\pmb{x}, t) \right\rbrace \rightarrow u(\pmb{x}, t)$ as $h \rightarrow 0$. Traditionally, $h$ and $N$ would be selected so that the Monte Carlo error is comparable to the bias, optimising the convergence rate. This approach is limited by the computational expense, which is proportional to $N\diagup h$. Alternatively, we can consider {\em debiasing} the estimator using such a numerical scheme, which is described in the remainder of this section. The reader is referred to \citet{Rhee2014} for debiasing expectations with respect to SDEs, and \citet{McLeish2011} for a more general discussion of debiasing Monte Carlo estimators.

Consider a monotonically decreasing series of step sizes, $\left\lbrace h_j \right\rbrace_{j=0}^\infty$, and note
\begin{align*}
u(\pmb{x}, t) & = E\left\lbrace\tilde{u}_{h_0} (\pmb{x}, t)\right\rbrace + \sum_{j=1}^{\infty} E\left\lbrace \left( \tilde{u}_{h_j} (\pmb{x}, t) - \tilde{u}_{h_{j-1}} (\pmb{x}, t) \right)\right\rbrace \\
& = \sum_{j=0}^{\infty} E\left\lbrace\tilde{\Delta}_j \right\rbrace,
\end{align*}
\noindent where we have denoted $\tilde{\Delta}_j = \tilde{u}_{h_j} (\pmb{x}, t) - \tilde{u}_{h_{j-1}} (\pmb{x}, t)$, and defined $\tilde{u}_{h_{-1}} (\pmb{x}, t) = 0$. 
It is possible to obtain an unbiased estimator of $u(\pmb{x}, t)$ by drawing a finite halting value, $H$, from some probability distribution $\mathcal{P}(H)$ that is independent of $\left\lbrace \tilde{u}_{h_j}(\pmb{x},t) \right\rbrace_{j=0}^\infty$, and for which $P(H \geq j) > 0$ for all $j \geq 0$. 
Define
\begin{equation*}
u^\dagger (\pmb{x},t) = \sum_{j=0}^H \omega_j \tilde{\Delta}_j,
\end{equation*}
\noindent where $\omega_j = 1 \diagup P(H \geq j)$, and note that
\begin{align*}
E \left\{ u^\dagger (\pmb{x},t) \right\} & = E \left\{ \sum_{j=0}^H \omega_j \tilde{\Delta}_j \right\} 
 = E \left\{ \sum_{j=0}^\infty \mathbb{I}_{H \geq j} \omega_j \tilde{\Delta}_j \right\} \\
& = \sum_{j=0}^\infty E \left\{\mathbb{I}_{H \geq j} \right\rbrace \omega_j E \left\{ \tilde{\Delta}_j \right\} 
 = \sum_{j=0}^\infty  E \left\{ \tilde{\Delta}_j \right\} 
 = u(\pmb{x}, t),
\end{align*}
\noindent where $\mathbb{I}_{j \leq H}$ is the indicator function
\begin{equation*}
\mathbb{I}_{H \geq j} =  \left\{ \begin{aligned} & 1 & \hspace{5pt} \mbox{if } H \geq j, \\
& 0 & \hspace{5pt} \mbox{otherwise}.\end{aligned}  \right.
\end{equation*}
\noindent If
\begin{equation*}
\sum_{j=1}^{\infty} \frac{ E \left\lbrace \left( \tilde{u}_{h_{j-1}} (\pmb{x}, t) - \tilde{u}_{h_{\infty}} (\pmb{x}, t) \right)^2 \right\rbrace } {P(H \geq j)} < \infty
\end{equation*}
\noindent then $u^\dagger (\pmb{x},t)$ is an unbiased estimator of $u(\pmb{x}, t)$ with 
\begin{multline}
\label{E:DVar}
E\left\lbrace u^\dagger (\pmb{x},t) ^2 \right\rbrace = \\ \sum_{j=0}^{\infty} \frac{ E \left\lbrace \left( \tilde{u}_{h_{j-1}} (\pmb{x}, t) - \tilde{u}_{h_{\infty}}(\pmb{x}, t) \right)^2 \right\rbrace - E \left\lbrace \left( \tilde{u}_{h_{j}} (\pmb{x}, t) - \tilde{u}_{h_{\infty}}(\pmb{x}, t) \right)^2 \right\rbrace } {P(H \geq j)}. 
\end{multline}
\noindent The necessary proof is given in Theorem 1 of \citet{Rhee2014}. Note that this proof requires that $\kappa(\pmb{x})$ be continuous. Whilst this will usually be the case for PDEs on infinite domains, it is not generally true of Dirichlet problems. In the following we set $N=1$ (each level of the approximation is estimated using a single sample path from the discretized SDE), and define $\left\lbrace h_j \right\rbrace_{j=0}^\infty$ through $h_0=t$, and the relation $h_{j+1} = h_j \diagup 2$. 
 
It is clear from (\ref{E:DVar}) that the efficiency of the debiasing approach depends strongly on the convergence rate of $\tilde{u}_h (\pmb{x},t)$, and hence how $\left\lbrace \tilde{u}_{h_j} (\pmb{x},t) \right\rbrace_{j=0}^{\infty}$ are jointly generated (termed the {\em coupling}). 
In \citet{Rhee2014} the view is taken that a single Brownian motion drives every level of the discretization. 
This can be achieved by either simulating the entire Brownian path at the finest discretisation level (conditional on the halting value), and then summing the components as necessary to evaluate the lower levels, or through the use of Brownian bridges from the Brownian path obtained at the initial level. 
Using this coupling, $E \left\lbrace \left( \tilde{u}_{h_{j}} (\pmb{x}, t) - \tilde{u}_{h_{\infty}} (\pmb{x}, t) \right)^2 \right\rbrace$ is $\mathcal{O}(2^{-jp})$, where $p$ is the strong order of the numerical discretization scheme. 
A lower variance estimator can be obtained by averaging over multiple, say $N^\dagger$, independent draws of $u^\dagger (\pmb{x},t)$,
\begin{equation*}
u^\ddagger (\pmb{x},t) = \sum_{i=1}^{N^\dagger} u^{\dagger (i)} (\pmb{x},t).
\end{equation*}





The expected computation time is proportional to $\sum_{j=0}^{\infty} 2^j P(H \geq j)$. 
A comparison with (\ref{E:DVar}) shows that the choice of distribution $\mathcal{P}(H)$ reflects a trade-off between lowering the variance of the estimator and computability. 
In \cite{Rhee2014} it is shown that if the strong order of the numerical scheme is greater than 0.5 then unbiased estimates can be obtained in finite expected computation time with square-root convergence rates.
For unbounded SDEs the Milstein approximation has strong order 1, giving an effective way of obtaining FKF estimates for PDEs on infinite domains.
However, when the SDE is bounded, such as in the Dirichlet problem, the strong order is reduced to 0.5.
This is a result of ignoring possible boundary interactions between discretely simulated points.
With the slower rate of convergence, a finite variance estimator can be obtained by selecting $\mathcal{P}(H)$ so that $P(H \geq j) = 2^{-jr}$ for some $r<2p$, but computability may be an issue.
\\

\noindent {\bf 5. Numerical Examples}

\noindent {\bf 5.1. 1D Example}

The first example is a 1D advection-diffusion model. Consider the following parabolic PDE, which has previously been used to study finite difference approaches \citep{Thongmoon2006},
\begin{equation*}
 \frac{\partial u(x,t)}{\partial t} + b \frac{\partial u(x,t)}{\partial x} = a \frac{\partial^2 u(x,t)}{\partial x^2},  \hspace{33pt} t \geq 0, \; x \in [0,1],
\end{equation*}
\noindent with initial and boundary conditions 
\begin{equation*}
\begin{aligned}
 u(x,0) & = 100x, &  x & \in [0,1],\\
 u(0,t)  & = 0, & t & \geq 0, \\
 u(1,t)  & = 100, & t & \geq 0.
\end{aligned}
\end{equation*}

The complementary SDE for this system is
\begin{equation*}
dX_s^x = -b\; ds + \sqrt{2a}\; dW_s,
\end{equation*}
\noindent which, when transformed using the Lamperti transform, becomes
\begin{align*}
dY_s^y & = -\frac{b}{\sqrt{2a}}\; ds + dW_s, \\
& = -\alpha\; ds + dW_s,
\end{align*}
\noindent where $y = \frac{x}{\sqrt{2a}}$. The boundaries for the transformed SDE are $0$ and $\frac{1}{\sqrt{2a}}$. Note that accepting proposed paths from this SDE does not require sampling from a Poisson process, owing to the fact that $\phi(v)$ is constant. Whilst this example is not fully demonstrative of the EA approach, it offers an analytical solution for comparison \citep{Thongmoon2006}.


A comparison of 95\% confidence intervals for the solution of the 1D advection-diffusion model obtained using CLT via the EA and debiasing approaches, over a range of parameter values at $t=5$, $x=0.9$, is given in Table~\ref{T:AD}, along with the average time taken to obtain each estimate. For the EA approach, the confidence interval is constructed from 1000 replicates of $\hat{u}(x,t)$, each obtained using 10 000 simulations, and for debiasing 1000 replicates of $u^\ddagger(x,t)$, each obtained using $N^\dagger = 10 000$. The halting value for the debiasing approach is drawn from a geometric distribution with parameter 0.45.

\begin{table}[h!]
\begin{center}
\caption{A comparison of 95\% confidence intervals of the solution to the advection-diffusion example at $t=5$, $x=0.9$, for a range of drift values.}
\label{T:AD}
\begin{tabular}{|c c c c c c c|}
\hline
a & b & True & \multicolumn{2}{c}{95\% Confidence Interval} & \multicolumn{2}{c|}{Average Time (s)} \\ 
 & & & EA & Debiasing & EA & Debiasing \\ 
\hline
0.01 & 0.1 & 56.13 & $56.11 \pm 2.3 \times 10^{-2}$ & $55.90 \pm 2.8 \times 10^{-1}$ & 0.140 & 0.436 \\ 
0.01 & 0.2 & 19.03 & $19.03 \pm 2.2 \times 10^{-2}$ & $18.94 \pm 6.5 \times 10^{-1}$ & 0.378 & 0.460 \\ 
0.01 & 0.3 & 5.223 & $5.227 \pm 1.3 \times 10^{-2}$ & $5.330 \pm 3.6 \times 10^{-1}$ & 1.248 & 0.216 \\ 
0.01 & 0.4 & 1.833 & $1.831 \pm 8.7 \times 10^{-3}$ & $2.141 \pm 3.7 \times 10^{-1}$ & 5.316 & 0.392 \\ 
\hline
\end{tabular}
\end{center}
\end{table}

Over the entire parameter range the EA approach has a lower work-variance product, showing that it is a more efficient approach than debiasing in this case. 
With regards to EA approaches, Brownian motion is a poor proposal when the P\'{e}clet number (proportional to $b \diagup a$) is large, leading to a large rejection rate. As can be seen in Table~\ref{T:AD}, as the P\'{e}clet number increases, the computational expense of obtaining the estimate increases, owing to the low acceptance rate of the proposed paths. This limitation is shared with finite difference and finite element methods, in which the resolution of the grid needs to be sufficiently large in order to avoid numerical instabilities when the P\'{e}clet number is large \cite{Onate2000}. 
The debiasing estimates are more variable in time taken to obtain an estimate, with the most expensive estimate taking 113 seconds in these experiments.
This demonstrates that extreme draws for the halting value can add a significant amount of computational expense

\noindent {\bf 5.2. 2D Example}

Now consider the following 2D PDE system,
\begin{equation*}
 \frac{\partial u(\pmb{x},t)}{\partial t} = \nabla k(\pmb{x}) \cdot \nabla u(\pmb{x},t) + \frac{1}{2} \nabla^2 u(\pmb{x},t), \hspace{33pt} t \geq 0, \; \pmb{x} \in [0,1] \times [0,1],
\end{equation*}
\noindent with initial and boundary conditions 
\begin{equation*}
\begin{aligned}
 u(\pmb{x},0) & = x_{(1)} x_{(2)}, &  x_{(1)} & \in [0,1], \; x_{(2)} \in [0,1], \\
 u(\pmb{x},t) & = 0, &  x_{(1)} & \in [0,1], \; x_{(2)}=0, \; t \geq 0, \\
 u(\pmb{x},t) & = x_{(1)}, &  x_{(1)} & \in [0,1], \; x_{(2)}=1, \; t \geq 0, \\
 u(\pmb{x},t) & = 0, &  x_{(1)} & = 0, \; x_{(2)} \in [0,1], \; t \geq 0, \\
 u(\pmb{x},t) & = x_{(2)}, &  x_{(1)} & =1, \; x_{(2)} \in [0,1], \; t \geq 0, \\
\end{aligned}
\end{equation*}
\noindent Here we choose
\begin{equation*}
k(\pmb{x}) = \exp\left(\frac{1}{2} x_{(1)} x_{(2)}\right).
\end{equation*}
\noindent The complementary SDE is
\begin{align*}
dX_{(1)} & = \frac{1}{2} X_{(2)} \exp \left(\frac{1}{2}X_{(1)} X_{(2)}\right) ds + dW_{(1)} \\
dX_{(2)} & = \frac{1}{2}X_{(1)} \exp \left(\frac{1}{2}X_{(1)} X_{(2)}\right) ds + dW_{(2)}.
\end{align*}
\noindent This reflects a challenging problem as the complementary SDE is nonlinear and we need exact first passage information for the boundary. Note that both of the major restrictions of our EA approach are met. The complementary SDE has unit volatility, and the potential function $A(\pmb{x})$ is given by $k(\pmb{x})$.

\begin{figure}[t]
\centering
\includegraphics[width=0.45\textwidth]{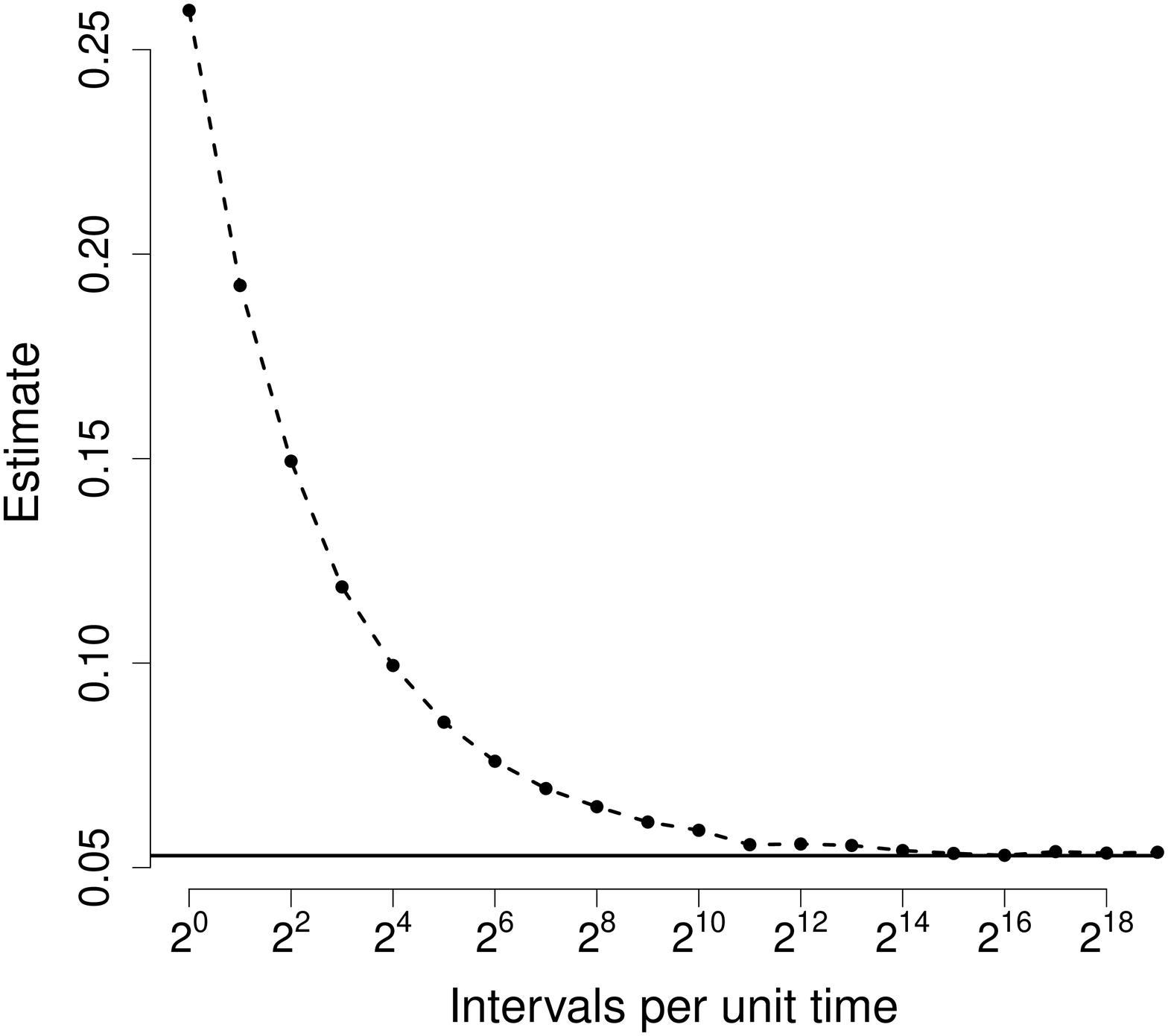}
\includegraphics[width=0.45\textwidth]{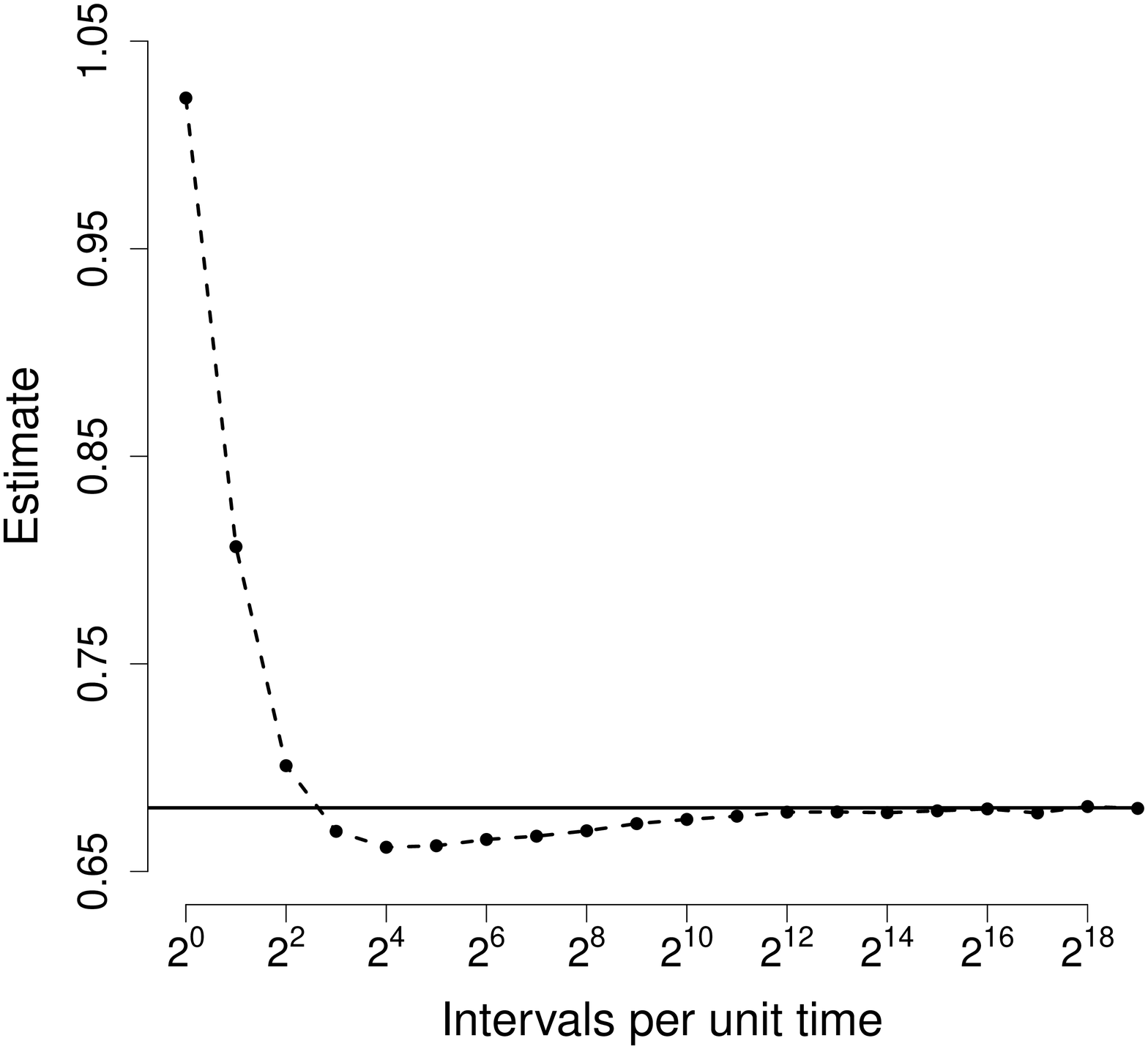}
\caption{Estimates of the FKF for the 2D PDE model at $t=2$, $\pmb{x}=c(0.2,0.2)^T$ (left), and $t=2$, $\pmb{x}=c(0.8,0.8)^T$ (right). In each case the Euler-Maruyama approximation (points) converges to the EA estimate (horizontal line).}
\label{F:2DAD}
\end{figure}

Since no analytical solution to this PDE exists, we compare an estimate obtained via EA with estimates obtained using the Euler-Murayama scheme with decreasing step sizes. It is known that in the limit of zero step size, the weak approximation error of the Euler-Murayama approach reduces to $0$ \citep{Gobet2000}. Hence, the Euler-Murayama estimates should converge to (within Monte Carlo error of) the EA estimate as the discretisation interval is reduced. The comparison is shown in Figure~\ref{F:2DAD} for $t=2$, $\pmb{x} = (0.2, 0.2)^T$ and $\pmb{x} = (0.8, 0.8)^T$, where $10^6$ simulations were used for each estimate, and we see that this is indeed the case. The EA estimates took 18 seconds and 13 seconds respectively to obtain, roughly the same as the Euler-Maruyama estimate with 1024 steps, and 512 steps respectively per unit interval. The discretisation error at this level of refinement is larger than the Monte Carlo error. These comparisons demonstrate that the EA approach is more computationally efficient in this example.

To compare the EA and debiasing approaches, we use the same setup as the 1D example. The 95\% confidence interval from the EA approach for $\pmb{x} = (0.2, 0.2)^T$ is $(5.29 \pm 0.01) \times 10^{-2}$, and from the debiasing approach is $(5.49 \pm 0.39) \times 10^{-2}$. The average time per estimate was 0.185 seconds for the EA approach, and 0.662 seconds for the debiasing approach. For $\pmb{x} = (0.8, 0.8)^T$ the 95\% confidence intervals were  $(6.81 \pm 0.001) \times 10^{-1}$ for EA and  $(6.84 \pm 0.21) \times 10^{-1}$ for debiasing, with average computation times per estimate of 0.126 s for EA, and 0.977 for debiasing. As with the 1D example, the EA approach gives lower variance estimators with a lower computational expense.\\

\noindent {\bf 6. Conclusions}

In this paper we have demonstrated how the FKF, combined with recent advancements in exact simulation of SDEs and debiasing methods, provides a way to obtain unbiased estimates of local solutions to a class of second order parabolic PDEs.
The practice of using finite difference or finite element methods to solve such problems leads to discretisation errors, and requires obtaining global solutions.
The FKF approach presented here therefore seems to be a more efficient approach in situations in which only local solutions are required. 
A well known example is the Bayesian inverse problem \citep{Stuart2010}, in which local solutions are required in order to obtain likelihood estimates given a set of data. 
The fact that the estimates presented here are unbiased means that it is possible to design exact Monte Carlo algorithms to sample from the posterior distributions, as shown in \citet{Herbei2014}. 
Even in situations where data are abundant, it is trivial to implement the FKF approach in parallel, limiting the computational expense to the single most expensive estimate. 

Exact simulation methods seem to outperform debiasing approaches, which suffer from both large estimator variance and computational expense when the domain of the PDE is bounded. 
However, current restrictions on exact algorithms, particularly in multiple dimensions, mean that they can not be used for a large number of interesting models. 
The reasons for this are twofold: in multiple dimensions it is not typically possible to transform an SDE to have unit diffusion, and the drift function must be of a suitable form to bound the Raydon-Nikod\'{y}m derivative.
It seems unlikely that these issues will be overcome in the current class of EA algorithms.
On the other hand, if improved convergence methods are developed for bounded SDEs, it is likely that debiasing will be more competitive with EA.
In \cite{Giles2008} it is empirically shown in one dimension that using stochastic interpolation techniques with discrete time approximations of SDEs speed up the convergence rates when the SDE is bounded.
To our knowledge no theoretical results have followed, and so we did not attempt this here.
 
A number of recent advances enable the use of FKF for a wider class of PDEs than those considered here.
In \citet{Pollock2014} exact algorithms are combined with $\epsilon$-strong simulation in order to estimate boundary crossing times to arbitrary precision. 
Although this approach is not unbiased, it can be applied to non-polytopal domains.
\citet{Jenkins2013} develops EA methodology for reflecting Brownian motion, enabling unbiased estimates of certain Neumann boundary problems.
\citet{Taylor2015} develops methodology for exact estimation of local times, extending the potential application to Neumann problems.
Finally, since the original formulation of the Feynman-Kac formulae there have been numerous extensions relating SDEs to other classes of PDEs, for example semilinear PDEs and nonlinear PDEs with forward-backwards SDEs \citep{Pardoux2014,Pham2014}, and higher order PDEs with iterated stochastic processes \citep{Thieullen2015}.
Simulating from the complementary SDE remains an essential step when using these extended formulas, and it is hoped that the work presented here will be built upon in these areas.



\vskip 14pt
\noindent {\large\bf Acknowledgements}

JC and MG were supported by EPSRC (ref: EP/L014165/1), and MP was supported by EPSRC (ref: EP/K014463/1).
\par

\markboth{\hfill{\footnotesize\rm Jake Carson, Murray Pollock and Mark Girolami} \hfill}
{\hfill {\footnotesize\rm Unbiased solutions of PDE Models} \hfill}

\bibhang=1.7pc
\bibsep=2pt
\fontsize{9}{14pt plus.8pt minus .6pt}\selectfont
\renewcommand\bibname

\end{document}